\documentclass[useAMS,usenatbib,preprint2]{aastex}
\usepackage{amsmath,amssymb}
\usepackage{delarray}
\usepackage{graphicx}

\newcommand{\Zsun}{{{\rm Z_{\odot}}}}

\newcommand{\diag}{\operatorname{diag}} 
\newcommand{\M}[1]{\mathbf{#1}}         

\newcommand{\fy}{{\rm f_{hot}}}
\newcommand{\hbr}{{\rm HBR}}

\newcommand{\Msun}{\ensuremath{\mathrm{M}_{\odot}}}

\newcommand{\nicefrac}[2]{\leavevmode\kern.1em
            \raise.5ex\hbox{\the\scriptfont0 #1}\kern-.1em
      /\kern-.15em\lower.25ex\hbox{\the\scriptfont0 #2}}

\begin{document}

\title{Fake star formation bursts: blue horizontal branch stars masquerade as young stars in optical integrated light spectroscopy.}
\shorttitle{Fake starbursts and BHB}

\author{P. Ocvirk\altaffilmark{1}}
\affil{Astrophysikalisches Institut Potsdam, an der Sternwarte 16, D-14482 POTSDAM, Germany}





\begin{abstract}
Model color magnitude diagrams of low-metallicity globular clusters usually show a deficit of hot evolved stars with respect to observations. We investigate quantitatively the impact of such modelling inaccuracies on the significance of star formation history reconstructions obtained from optical integrated spectra. To do so, we analyse the sample of spectra of galactic globular clusters of Schiavon et al. with STECKMAP (Ocvirk et al.) and the stellar population models Vazdekis et al. and Bruzual \& Charlot, and focus on the reconstructed stellar age distributions.
Firstly, we show that background/foreground contamination correlates with E(B-V), which allows us to define a clean subsample of uncontaminated GCs, on the basis of a E(B-V) filtering.

We then identify a ``confusion zone'' where fake young bursts of star formation pop up in the star formation history although the observed population is genuinely old. These artifacts appear for 70-100\% of cases depending on the population model used, and contribute up to 12\% of the light in the optical.
Their correlation with the horizontal branch ratio indicates that the confusion is driven by HB morphology: red horizontal branch clusters are well fitted by old stellar population models while those with a blue HB require an additional hot component. The confusion zone extends over {\rm [Fe/H]}$=[-2,-1.2]$, although we lack the data to probe extreme high and low metallicity regimes. 
As a consequence, any young starburst superimposed on an old stellar population in this metallicity range could be regarded as a modeling artifact, if it weighs less than 12\% of the optical light, and if no emission lines typical of an HII region are present.
This work also provides a practical method for constraining horizontal branch morphology from high signal to noise integrated light spectroscopy in the optical. This will allow post-AGB evolution studies in a range of environments and at distances where resolving stellar populations is impossible with current and planned telescopes.

\end{abstract}


%




\section{Introduction}

Spectroscopy is a major tool for the study of galaxy formation and evolution. The analysis of the spectrum of galaxies provides constraints on their star formation activity, past and present. However the accuracy of these constraints  depends on the quality of the data, but also critically on the quality of the models used to interpret these spectra, as shown for example in \cite{koleva08}. 
Since spectral models of stellar populations rely on stellar evolution models, it is vital the latter models be able to account for most of the observed stellar evolution stages, or at least the most luminous ones (e.g. turnoff, red giant branch and horizontal branch stars).
While thermally pulsing asymptotic giant branch (TP-AGB hereafter) stars are fundamental for NIR spectral modelling, hot evolved stars are crucial in the UV for intermediate and old populations. The latter have been proposed as the origin of the UV-upturn of elliptical galaxies \citep{han07,Yi2008}. In this paper, we will focus on these hot  horizontal branch (hereafter HB) stars. They are fairly luminous for their mass: while weighing only about 0.5 $\Msun$, they have a typical absolute magnitude M$_{V}$=-0.7 \citep{allens}. Their He-burning core tends to be hotter with decreasing metallicity, making the HB of low-Z galactic globular clusters bluer \citep{lee94}, although the morphology of the HB also depends on a second parameter, so far elusive. 

Accurate modelling of the HB morphology of GCs is a difficult problem. 
For instance, it is strongly sensitive to mass-loss on the RGB \citep{dotter08}. Moreover, 
accurate fitting of blue HB in GCs usually requires a fine tuning of He abundance dispersion, as shown in NGC2808 by \cite{lee05}. 
This clashes with the general treatment of spectral synthesis for galaxies, where a single stellar population (hereafter SSP) is chemically homogeneous. In effect, the parameter space allowed by popular full spectrum SSP models (as opposed to those predicting indices) such as \cite{BC03} or \cite{PEG-HR} is strictly (age,metallicity), and the abundances are solar-scaled, with no intrinsical dispersion. With only these parameters, it is expected that the isochrones adopted in these SSP models are not appropriate for the whole range of HB morphology.
Indeed, Fig. 7 of \cite{BC03} shows that the 'Padova 1994' and 'Padova 2000' tracks do not fit well the color-magnitude diagrams (hereafter CMD) of GCs with blue HB. In particular the HB of the models does not extend enough towards the blue. These very tracks are nonetheless used for the computation of spectral SSP models.
What can be the impact of this lack of hot evolved stars when interpreting the spectra of stellar populations ? Their effect can be readily seen in the integrated colors of GCs \citep{smith07}.
Moreover, recent works indicate that space observations in the GALEX NUV and FUV are required to differentiate between blue HB stars and young hot stars \citep{kaviraj07,schawinski07kav}. However, although GALEX has been successful at securing data for a fair fraction of the sky, a large corpus of past, present and ongoing extragalactic studies still relie on optical spectra only (e.g. \cite{moped01,panter1,heavensnature,vespa07,vespa09,koleva09,michielsen08}).

It is now well known that HB morphology makes age-dating stellar populations from optical spectra only troublesome, using indices \citep{beasley2000,puzia05}, but also using full spectrum fitting techniques \citep{koleva08,sharina09}: the age inferred spectroscopically can be several Gyr younger than that measured from the turnoff.
The effect on star formation history (hereafter SFH) reconstructions could be even more dramatic:
do the missing hot stars show up as an additional fake burst as suggested in \cite{coelho09}, or do they simply shift the whole age distribution to younger ages ? At the moment, no studies have been published specifically on the effect of HB morphology on the significance of SFH reconstructions from full spectrum fitting in the optical.

In this paper, we propose to investigate quantitatively this very issue. To do so, we first need a sample of objects with well-known, simple star formation histories, and a variety of well constrained HB morphologies. From this point of view, GCs are advantageous since they are mostly believed to arise from a single star formation episode, and display a whole range of HB morphologies. We will focus on the library of integrated spectra of galactic GCs of \cite{schiavon05} with STECKMAP \citep{STECMAP,STECKMAP}, a Bayesian method for reconstructing the star formation history of galaxies using full spectrum fitting, and a collection of popular single stellar population (SSP) models based on MILES \citep{MILES} and also \cite{BC03}. 
For each GC, the CMD is available from the literature, so that we can relate HB morphology to  possible departures of the SFH reconstructions from a unique star formation burst.
We recall that the aim of the paper is not to propose a new spectral model for the sample of GCs: it is well established that interpreting HB morphology in globular clusters requires more parameters than the simple SSP population models used here \citep{bedin04,ferraro04,norris04}. Instead, our goal is to understand how the unknown HB morphology in galaxies may complicate the interpretation of their spectra.
That's why our setup reproduces the prevailing regime in which most extragalactic stellar populations studies in optical integrated light spectroscopy are carried out and therefore uses SSPs or combinations of SSPs.
The plan of the paper is as follows: Sec. \ref{s:methodology} presents the data, the SSP models and the method, and Sec. \ref{s:results} presents the recovered stellar age distributions and their dependence on HB morphology. In Sec. \ref{s:discussion}, we discuss the implications of our results for spectroscopic extragalactic studies.






\section{Methodology}
\label{s:methodology}
\subsection{Data}
We consider the sample of integrated spectra of galactic globular clusters published by \cite{schiavon05}\footnote{http://www.noao.edu/ggclib/description.html}.
It consists of 40 Galactic globular clusters, obtained with the Blanco 4 m telescope and the R-C spectrograph at the Cerro Tololo Inter-American Observatory. The spectra cover the range $\approx3350-6430$ {\AA} with $\approx3.1$ {\AA} (FWHM) resolution. The spectroscopic observations and data reduction were designed to integrate the full projected area within the cluster core radii in order to properly sample the light from stars in all relevant evolutionary stages. The signal to noise values of the flux-calibrated spectra range from 50 to 240 per {\AA} at 4000 {\AA} and from 125 to 500 per {\AA} at 5000 {\AA}. The selected targets span a wide range of cluster parameters, including metallicity, horizontal-branch morphology, Galactic coordinates, Galactocentric distance, and concentration.
We dropped NGC6569, NGC6388 and NGC6441, which have rather peculiar HB morphology.
We are then left with 37 GCs. The parameters relevant to this study are listed in Tab. 1, and are taken from Table 1 of \cite{schiavon05}. 
In order to characterize HB morphology, we use the Horizontal Branch Ratio (HBR) indicator, defined as in \cite{lee94}: \begin{equation}
{\hbr = \rm (B-R)/(B+V+R) \, ,}
\end{equation}
 where B, V, R are the number of blue, variable and red HB stars, respectively. GCs with a blue HB have $\hbr \approx 1$, while GCs with a red HB have $\hbr \approx -1$.
For NGC5946 and NGC6284, we assign $\hbr=1$ (no value is provided in \cite{schiavon05}).
Since the GCs we study here have masses of the order $10^{5}-10^{6} \Msun$ (see \cite{chernoff89} for instance for mass determinations), and the scanned area encloses a large fraction of the mass, we consider that our results should not be affected by stochastic fluctuations related to the finiteness of their stellar populations. According to \cite{lancon2000}, the flux fluctuations for our sample should be smaller than a few percents.
\begin{figure*}[t]
\includegraphics[width=0.33\linewidth]{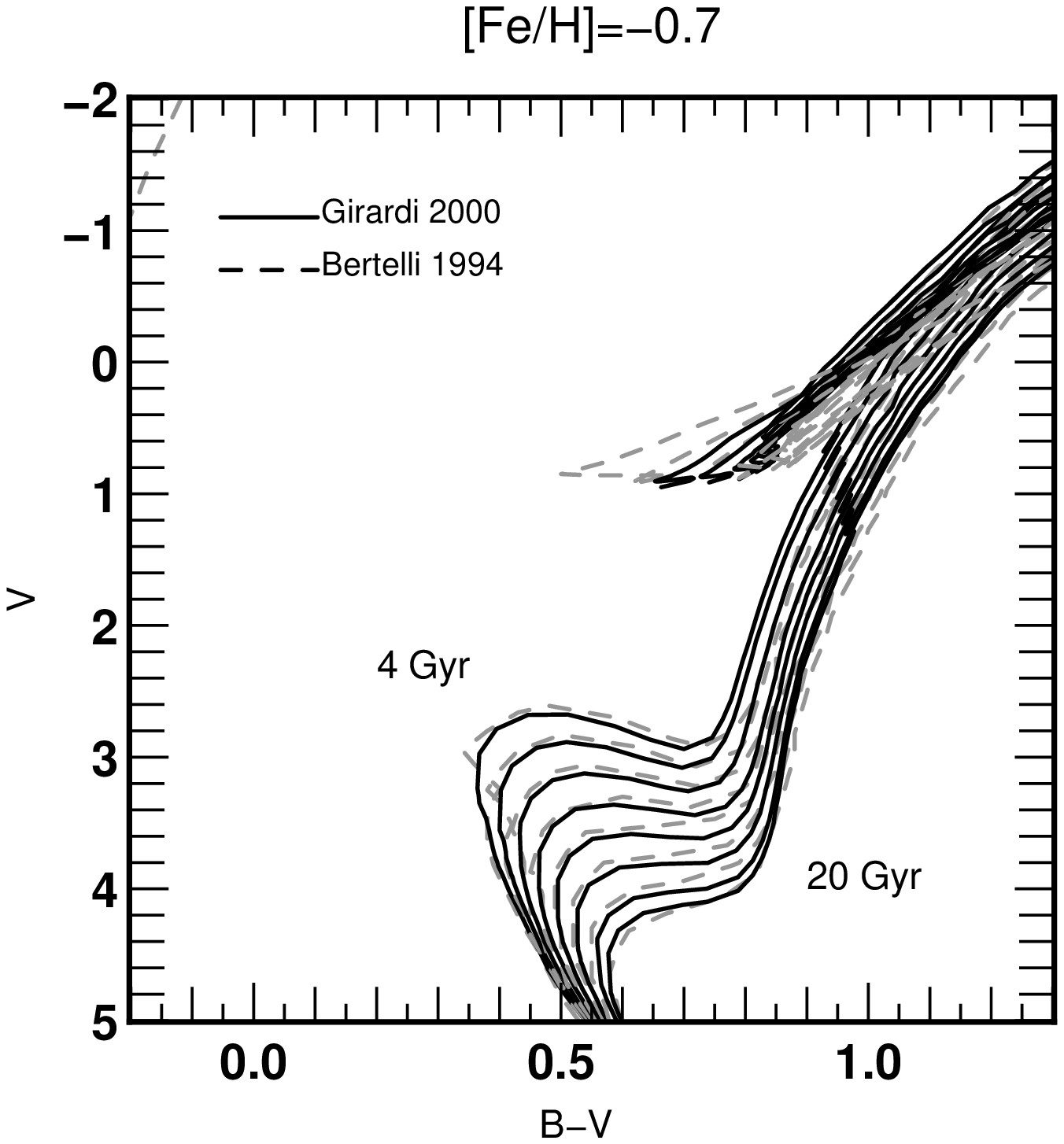}
\includegraphics[width=0.33\linewidth]{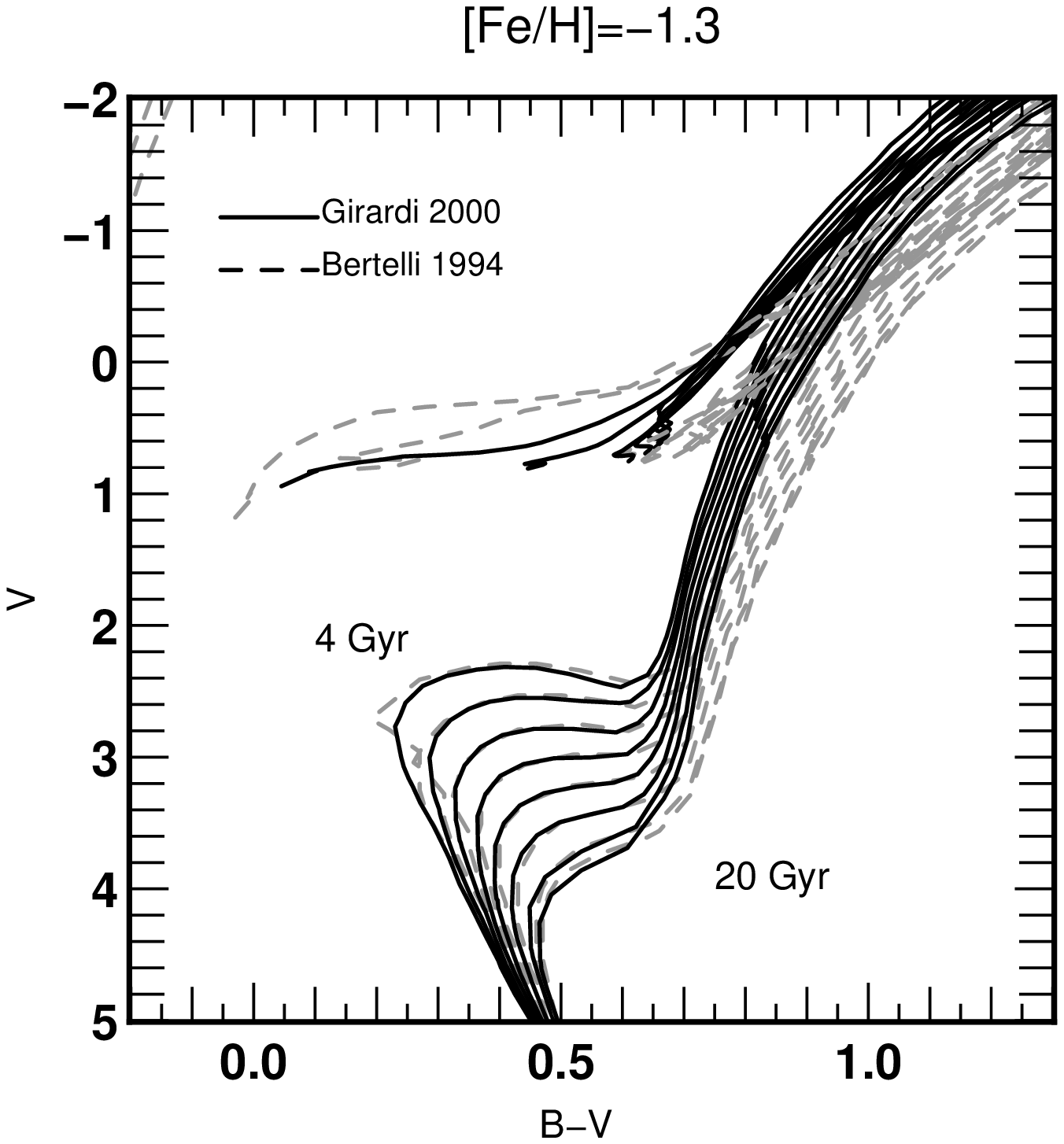}
\includegraphics[width=0.33\linewidth]{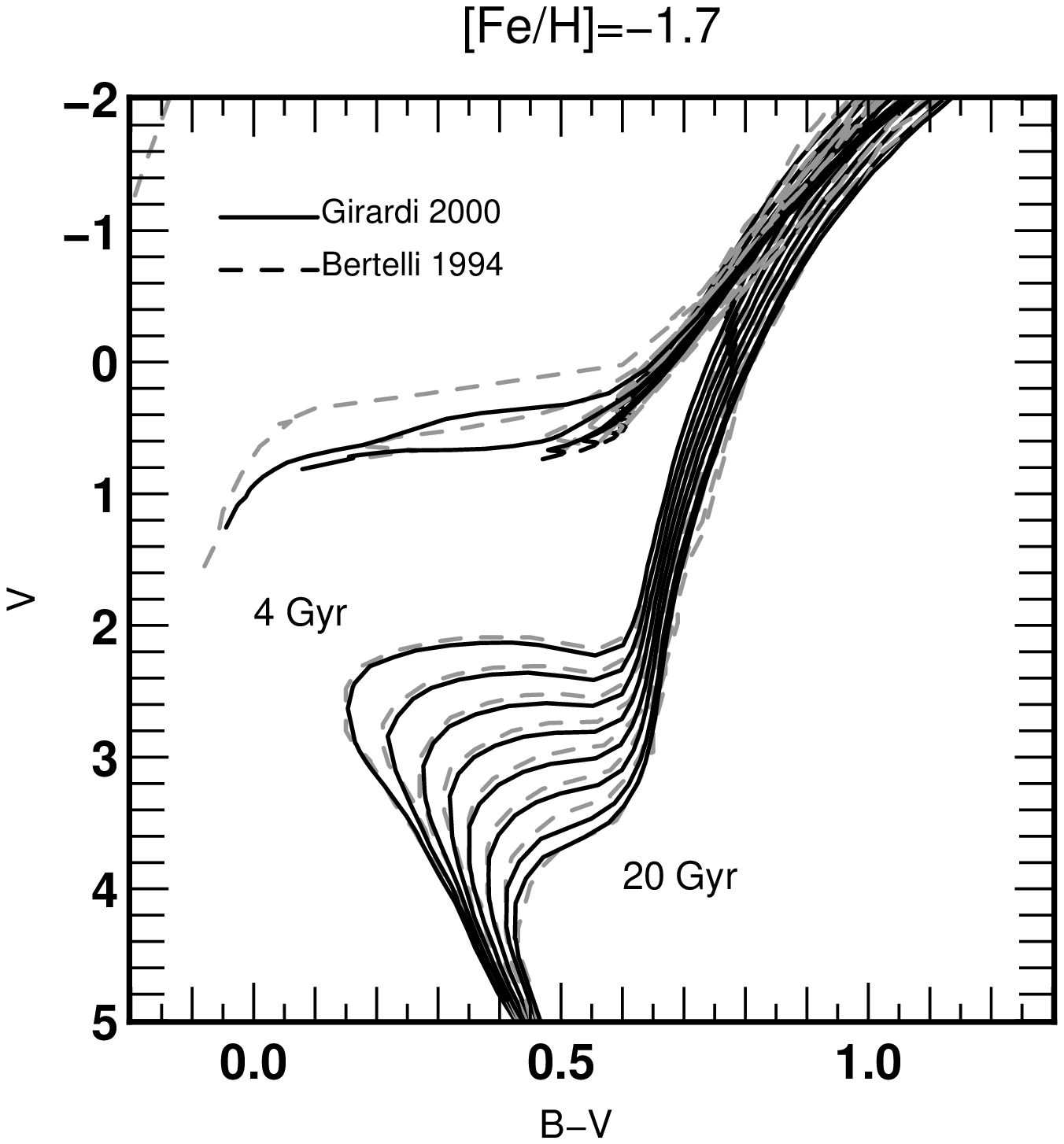}

\caption{Color-magnitude diagrams of the isochrones used for the Vazdekis et al. models (\cite{girardi00}, black solid line) and BC03 (\cite{bertelli94}, grey dashed line) for 3 metallicities {\rm [Fe/H]}=[-0.7,-1.3,-1.7]. For each panel, the sequence spans the [4-20] Gyr age range by logarithmic increments of 0.1 dex.}
\label{f:isocs}
\end{figure*}

\subsection{Models}
We consider two popular SSP models.
The first one (in preparation, hereafter Vazdekis et al.\footnote{http://www.iac.es/galeria/vazdekis/vazdekis\_models\_ssp\_seds.html}) is based on the MILES stellar library \citep{MILES} and the isochrones from \cite{girardi00}. The other one is \cite{BC03} (hereafter BC03), based on the STELIB stellar library \citep{stelib}, and using the ``Padova 1994'' tracks \citep{bertelli94}. The corresponding isochrones for [Fe/H]$=[-0.7,-1.3,-1.7]$ are shown in Fig. \ref{f:isocs}. Note that throughout the paper we will use the notation [Fe/H]$=\log_{10}({\rm Z}/{ \Zsun})$, and $\Zsun=0.0189$ \citep{anders89}.

The isochrones are very similar up to the beginning of the RGB. In the latter phase, the Girardi 2000 set is bluer, especially for {\rm [Fe/H]}=-1.3. Then both sets develop a horizontal branch which extends further to the blue with increasing age and decreasing metallicity. 

The spectral models we use thus include the HB contribution by construction.  However, although the predicted {\em position} of the BHB in the CMD is rather satisfactory in these isochrone sets, the predicted {\em{number}} of stars in this phase is very small. For instance, a synthetic CMD obtained with IAC-STAR\footnote{http://iac-star.iac.es/iac-star/} in the default setup using the \cite{girardi00} isochrones for a {\rm [Fe/H]}=-1.7, 17 Gyr old population, produces a horizontal branch ratio $\hbr=0.2$, while most GCs of our sample have $\hbr=1$ at the same metallicity and an age of about 10 Gyr.
In effect, we checked carefully the spectral evolution of the low-Z models from 10 to 20 Gyr in the optical and they show very little sign of the increasing BHB contribution with age: for the Vazdekis et al. models, a very slight bluening and deepening of the Balmer lines (about 1\%) can be seen at these extreme ages, however for the BC03 models we see nothing at all.
In any case the contribution of the BHB to the optical spectrum even in 20 Gyr models is much smaller than the difference in GCs spectra induced by differences in HB morphology for clusters of similar metallicity, as shown for instance by Fig. 3 of \cite{schiavon05}. So even if the BHB is present {\em in the isochrones}, it is legitimate to say that there is almost no BHB contribution {\em to the spectra} in the optical within our set of SSP models.

Both models cover equally well the wavelength range of the data.
We consider the metallicity set ${\rm [Fe/H]}=[-1.7,-0.7,-0.4,0,0.4]$.
Although BC03 covers a wider age range, we restrict ourselves to SSP sequences of age=$[10^8-2. 10^{10}]$ yr for both models, divided in 40 age bins. We include these extreme ages even though they are larger than the WMAP5 age of the universe of 13.7 Gyr \citep{WMAP5}, because, as shown by Fig. \ref{f:isocs}, really blue horizontal branches develop only at ages larger than 12 Gyr in both sets of isochrones.
We did not use PEGASE-HR \citep{PEG-HR} coupled with the ELODIE 3.1 library \citep{elodie31}, although it offers the highest spectral resolution (R=10000), because it does not cover the $\lambda < 3900$ {\AA} domain.

\begin{figure*}[t]
\rotatebox{270}{\includegraphics[width=0.38\linewidth]{{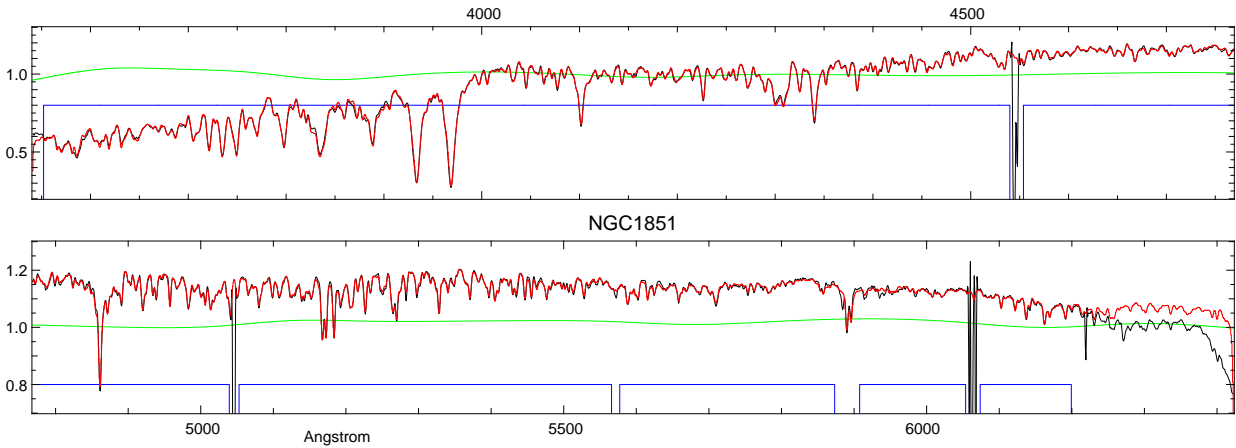}}}
\caption{Spectrum of NGC1851 ({\em black}) and best fitting model using STECKMAP and the Vazdekis et al. SSPs ({\em red}). The green line represents the non-parametric transmission curve as explained in Sec. \ref{s:steckmap}. Bad pixels and some residuals of sky subtraction have been masked, as well the very blue and the red end of the spectrum beyond 6200 {\AA}.}
\label{f:spectra}
\end{figure*}

\subsection{Interpretation software: STECKMAP}
\label{s:steckmap}
In order to analyse the observed spectra with SSP models as a reference, we use the interpretation code STECKMAP \citep{STECMAP,STECKMAP}. The SSP models are used in a flux-normalized form (i.e. the average flux of every SSP is 1), rather than mass-normalized, as in \cite{STECMAP} and \cite{STECKMAP}. In this setting, STECKMAP returns the most likely stellar age distribution (hereafter SAD) corresponding to the analysed spectrum, i.e. the luminous weight of each model component as a function of age, thereby providing an analog of the star formation history\footnote{The star formation history can be obtained by multiplying the stellar age distribution by the mass-to-light ratio of each bin divided by the time spanned by each time bin}, giving insight into the luminous balance between various stellar components seen in the spectrum.
One important aspect of STECKMAP is that it is a regularized method, i.e. the various unknowns (here SAD, age-Z relation, and line-of-sight velocity distribution) are forced to be smooth in order to avoid exaggeratedly oscillating solutions. The smoothing kernel is laplacian for the SAD (i.e. $\M{L}=\M{D}_2$ using the notations of \cite{STECKMAP}), and we chose $\mu_{\M{X}}=1$, which gives robust results. 

Besides, we require the age-Z relation to be flat (i.e. mono-metallic population), and implement this constraint by using a finite difference operator $\M{L}=\M{D}_1=\diag_{2}[-1,1]$ and a large smoothing parameter $\mu_{Z}=10^5$. 
The smoothing kernel for the line-of-sight velocity distribution is laplacian as for the SAD, and the smoothing parameter was set to $\mu=10^2$, which produces a smooth gaussian LOSVD for the whole sample. 
In order to deal with possible flux calibration errors, we mutliply the model by a smooth non-parametric transmission curve, representing the instrumental response multiplied by the interstellar extinction. This curve has 30 nodes spread uniformally along the wavelength range, and the transmission curve is obtained by spline-interpolating between the nodes. The latter are treated as additional parameters and adjusted during the minimization procedure. This continuum matching technique is similar in essence to the multiplicative polynomial used by NBursts \citep{NBURSTS}, and will be detailed in a forthcoming paper.

Finally, the first guess for the solution consists of a flat, constant SAD. However, in some occasions, we used a family of SADs with a unique gaussian burst centered on a random age to probe the existence of local minima (see Sec. \ref{s:youngfBHB}).

\begin{figure*}[t]
\includegraphics[width=0.5\linewidth]{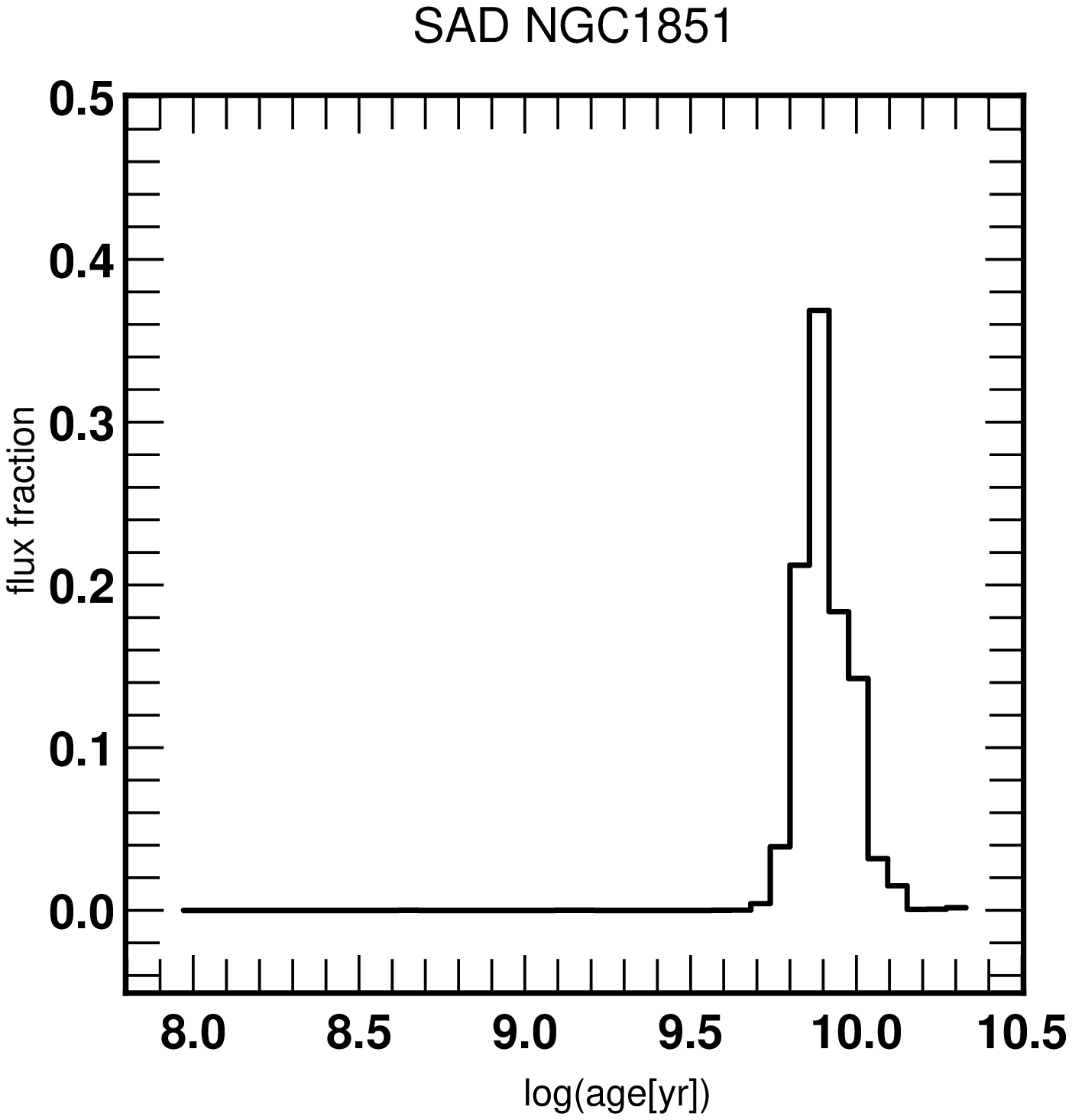}
\includegraphics[width=0.5\linewidth]{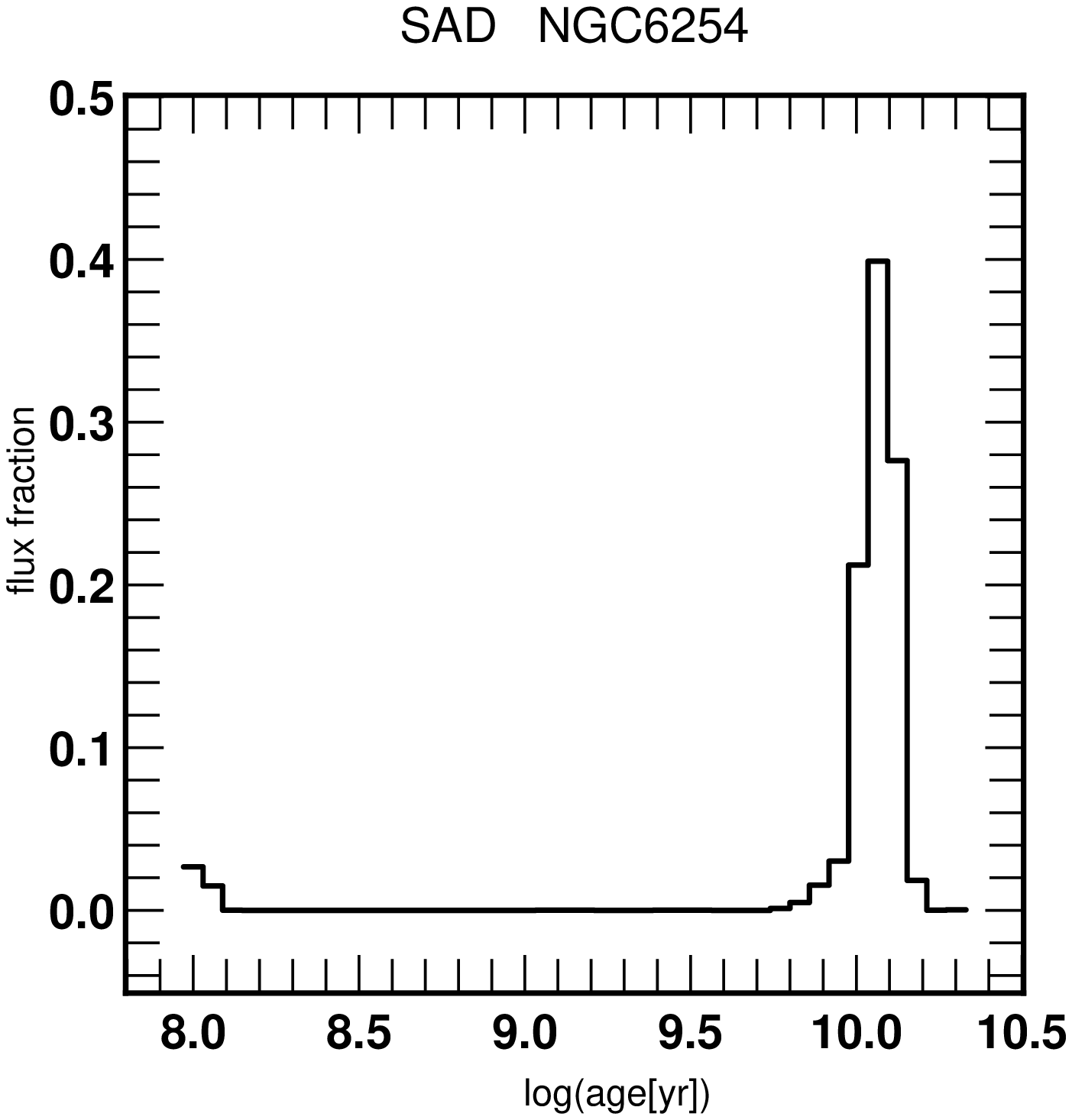}
\caption{Stellar age distributions of NGC1851 and NGC6218 using the Vazdekis et al. models.}
\label{f:2SADs}
\end{figure*}

\section{Results}
\label{s:results}

We now present the results obtained by applying STECKMAP with the described setup to the sample of GCs. A very good fit to the data is obtained in all cases, as shown for NGC1851 in Fig. \ref{f:spectra}.

\subsection{2 families of stellar age distributions}
\label{s:sads}
We find that the recovered SADs fall in two classes, as illustrated by Fig. \ref{f:2SADs}:
\begin{itemize}
\item{type I: genuinely old SAD, with a single bump consisting only of $>10^{9.5}$ yr (left panel).}
\item{type II: two-component SAD, with most of the light provided by an old population, and a small light fraction (up to 12\%) coming from the youngest bins (right panel).}
\end{itemize}

While the first family of SADs is what one would actually expect for GCs, the second one comes as a surprise, since GCs are usually expected to consist of stars formed in one single short star formation event. We already pointed out that hot evolved stars such as BHB stars, if unaccounted for in the SSP models, could show up in the spectra by producing abnormally deep blue Balmer lines. In such cases, the model SSPs will be redder and display weaker Balmer lines than the observed spectrum. As a consequence, our interpretation code STECKMAP indicates that the best fitting combination of the SSP models provided consists of an old SSP plus a very young SSP that mimics the HB stars unaccounted for by the model isochrones.

We now test the relevance of this interpretation of type II SADs by investigating quantitatively their connection with HB morphology.



\subsection{Hot fraction and horizontal branch morphology} 
\label{s:youngfBHB}

\begin{figure}
\includegraphics[width=1.0\linewidth]{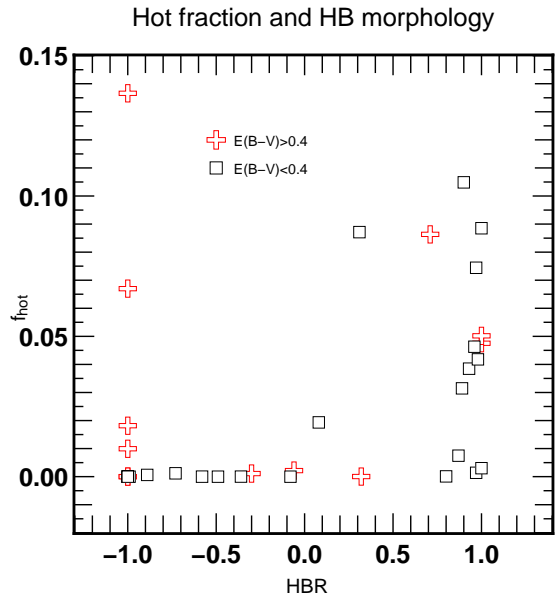}
\caption{Hot fraction computed from the stellar age distributions versus horizontal branch ratio (HBR), using the Vazdekis et al. models. {\em Crosses:} GCs with E(B-V)$\ge$0.4, {\em squares:} GCs with E(B-V)$\le$0.4}
\label{f:HBRwhole}
\end{figure}
We define the hot fraction $\fy$ as the sum of the luminous weights of the young (age $<10^{8.5}$ years) components.
We recall here that the SAD is normalized (the sum of all weights is 1).
In Fig. \ref{f:HBRwhole} we plot the hot fraction $\fy$ versus the HBR for the whole sample. 
About half of the clusters have $\fy = 0$, which was our initial expectation. 
Moreover, there seems to be a general trend of increasing $\fy$ with HBR.
About 10 of the GCs have $\hbr \approx 1$ {\em} and $\fy > 0$, which supports our interpretation for the type II SADs, as explained in Sec. \ref{s:sads}. 
There are however outliers with respect to this simple picture: 4 GCs with red HB have $\fy > 0$, and another 3-4 have a blue HB but $\fy \approx 0$. We checked that these were not secondary minima by using a family of single bursts populations centered on random ages as first guess. They all led to the same solution.

\subsubsection{Background/foreground contamination}
\begin{figure*}[t]
\includegraphics[width=0.5\linewidth]{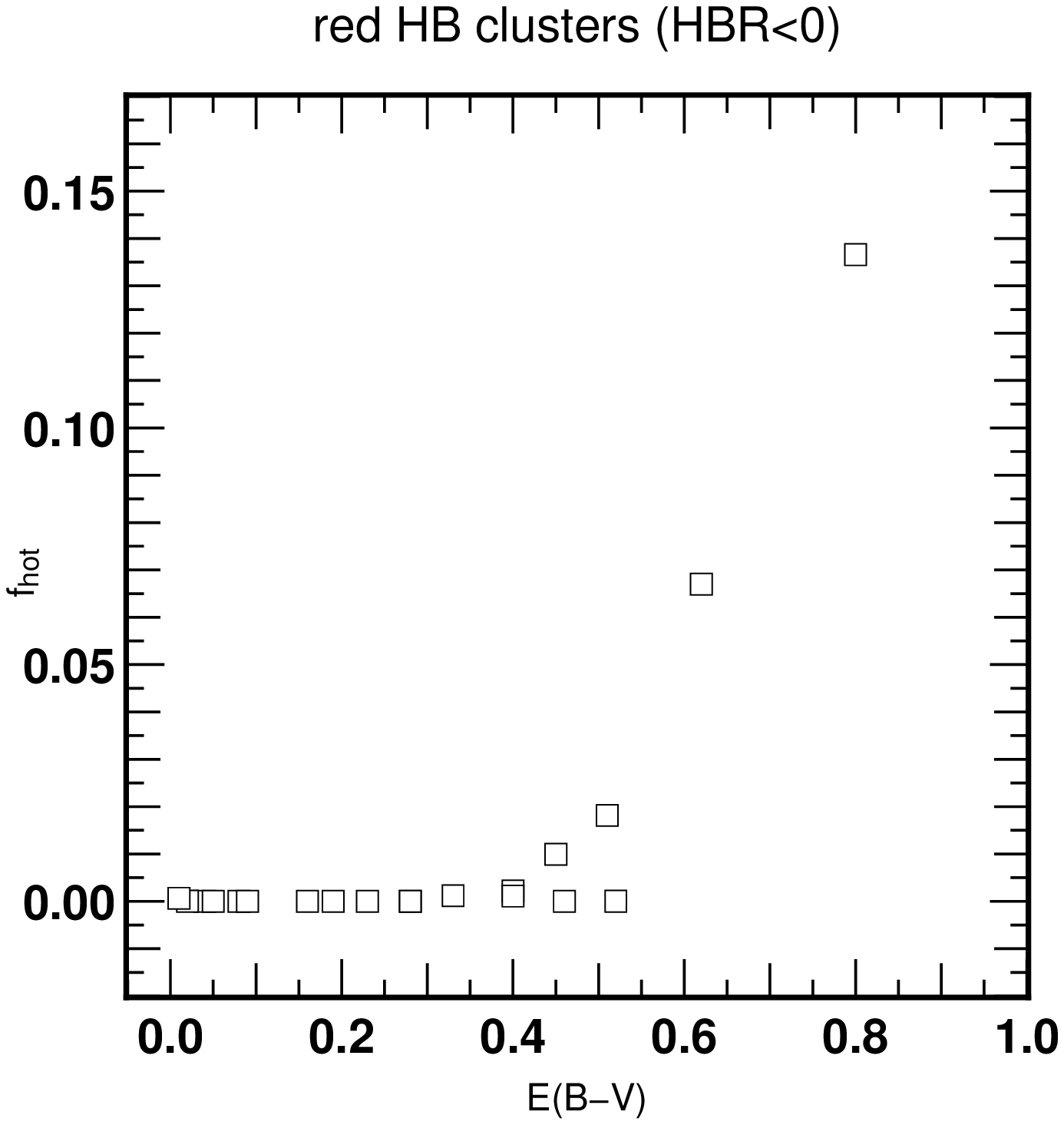}
\includegraphics[width=0.5\linewidth]{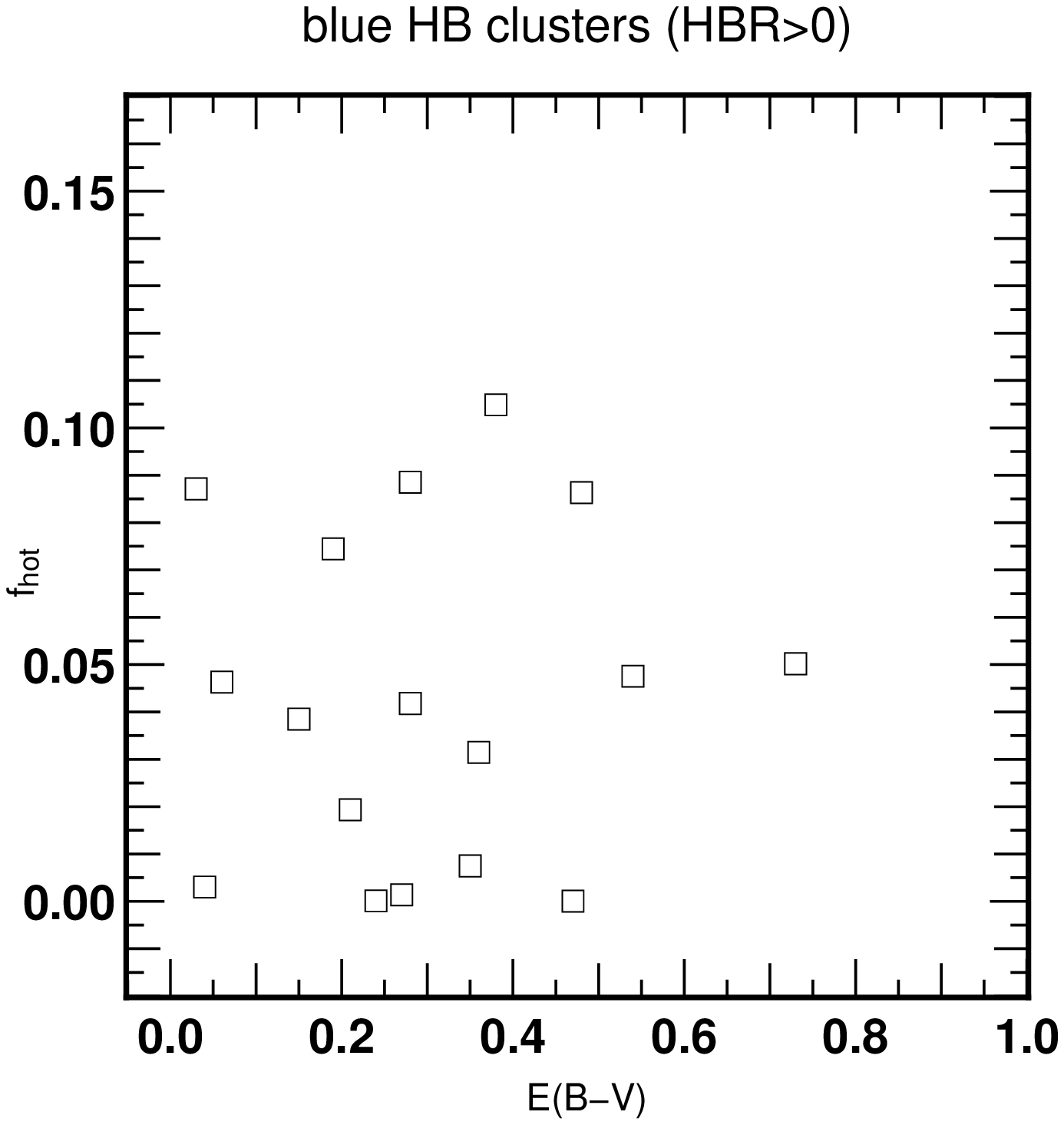}
\caption{Hot fraction as a function of color excess, using the Vazdekis et al. models. {\em Left:} red HB clusters, {\em right:} blue HB clusters.}
\label{f:contam}
\end{figure*}

The two objects in the top left part of Fig. \ref{f:HBRwhole} are NGC6553 ad NGC6528, which are located in the direction of the galactic center. 
Because of their location, they are definitely the most likely to be contaminated. 


Since we are interested in the effect of HB morphology on the spectra rather than contamination, we wish to work on a sample free from foreground/background contamination.
One could define a subsample by removing the GCs too close to the galactic center direction. However, this does not provide a good filtering, since the likelihood of a contamination depends on the integrated stellar density inside the scanned volume and the exposure time, and hence also on the distance to the GC. 
We thus investigated the use of alternative indicators of contamination.



Red HB morphology is usually well reproduced by the tracks used for stellar populations synthesis (see Fig. 7 of \cite{BC03}): unlike for blue HB GCs, old SSP models suffice to reproduce the spectra of red HB GCs. Because of this superior agreement with models, we can use them as calibrators: any deviation from type I SAD in a red HB GC should be caused by background/foreground contamination rather than inaccurate evolutionary models. Therefore, for these objects, $\fy$ should trace contamination.
One could object that the large $\fy$ found could also be a consequence of a strong contribution from blue stragglers (hereafter BS). Indeed, \cite{zoccali01} shows a clear ``blue plume'' in the CMD of NGC6553. However, their kinematical cleaning shows that only half of the stars of the plume are indeed cluster members and therefore genuine BS. The other half of the plume consists of young disk stars in the foreground, and is hotter and more luminous than the BSs. 
Hence, even if BSs certainly partially contribute to the blue light excess, the latter is dominated by foreground disk contamination. The high $\fy$ found for NGC6553 thus effectively traces contamination rather than abnormal BSs frequency.

The left panel of Fig. \ref{f:contam} shows $\fy$ as a function of color excess for GCs with $\hbr<0$. Indeed, most of the red HB GCs are well fitted by genuinely old SSPs, without the need for any additional hot component. However, for E(B-V)$>0.4$, $\fy$ increases sharply, reaching a maximum for the 2 galactic center GCs NGC6553 and NGC6528.
This correlation between $\fy$ and extinction lends support to our interpretation.
In effect, a correlation between background/foreground contamination and extinction should naturally arise from the two following points:
\begin{itemize}
\item{if dust and stellar density correlates, then E(B-V) increases with the integrated stellar density along the line of sight to the GC.}
\item{young hot contaminating stars are more likely to live in dusty environments.}
\end{itemize}
It seems therefore that the color excess E(B-V) is a good proxy for the likelihood of foreground/background contamination, and the left panel of Fig. \ref{f:contam} tells us that GCs with E(B-V)$<0.4$ should be relatively free from contamination.
In order to clean our sample from contaminated spectra, we will drop the high-extinction GCs, and keep only those with E(B-V)$<0.4$.
The right panel of Fig. \ref{f:contam} shows that, within this new definition of the sample, there is no correlation between E(B-V) and $\fy$ for the blue HB GCs either, and the points are randomly distributed. It also shows that the E(B-V) filtering only removes 4 BHB GCs, and leaves us with a large enough sample to carry on with our investigation.



\subsubsection{The $\fy-\hbr$ correlation}
\begin{figure*}[t]
\includegraphics[width=0.5\linewidth]{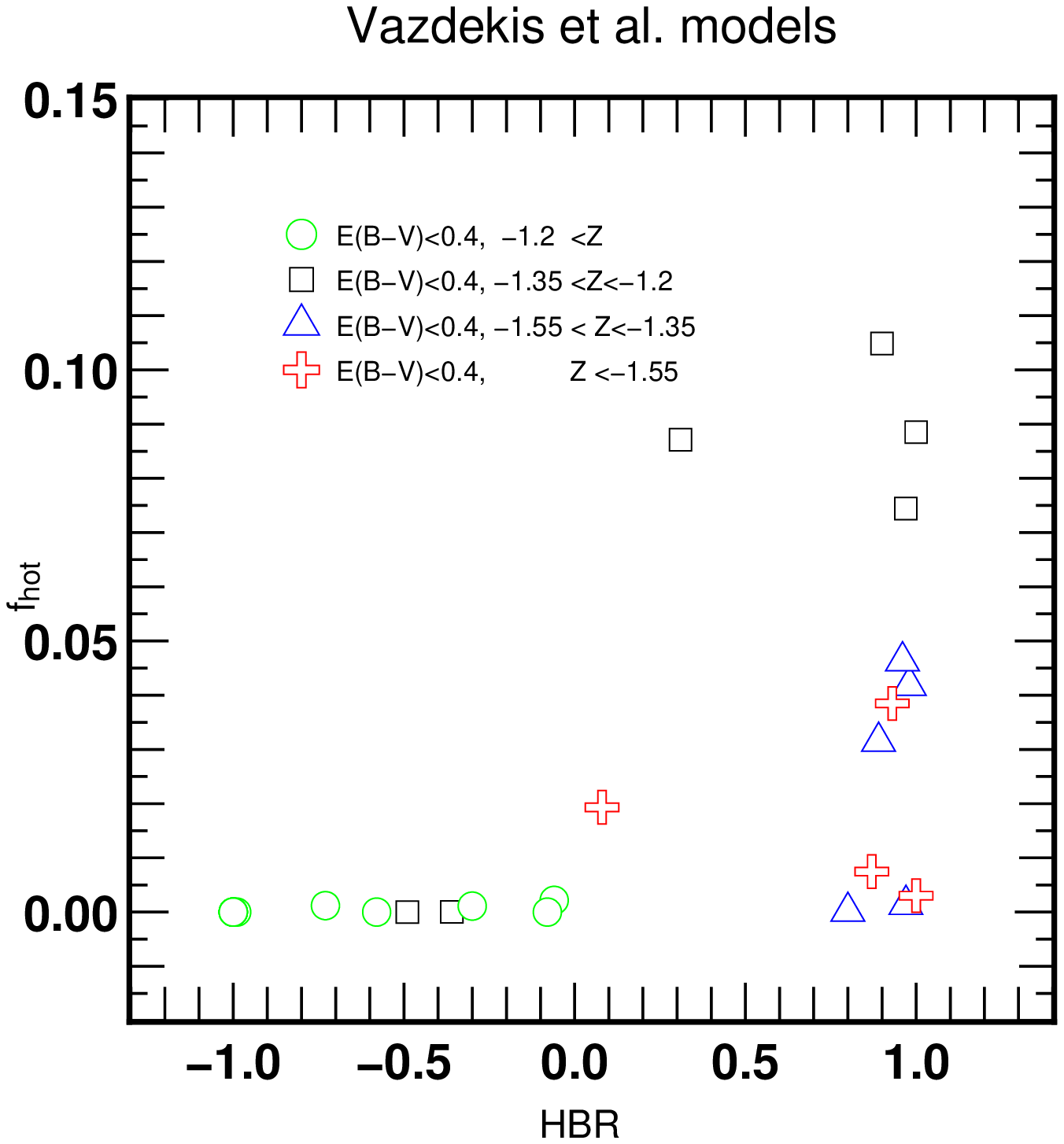}
\includegraphics[width=0.5\linewidth]{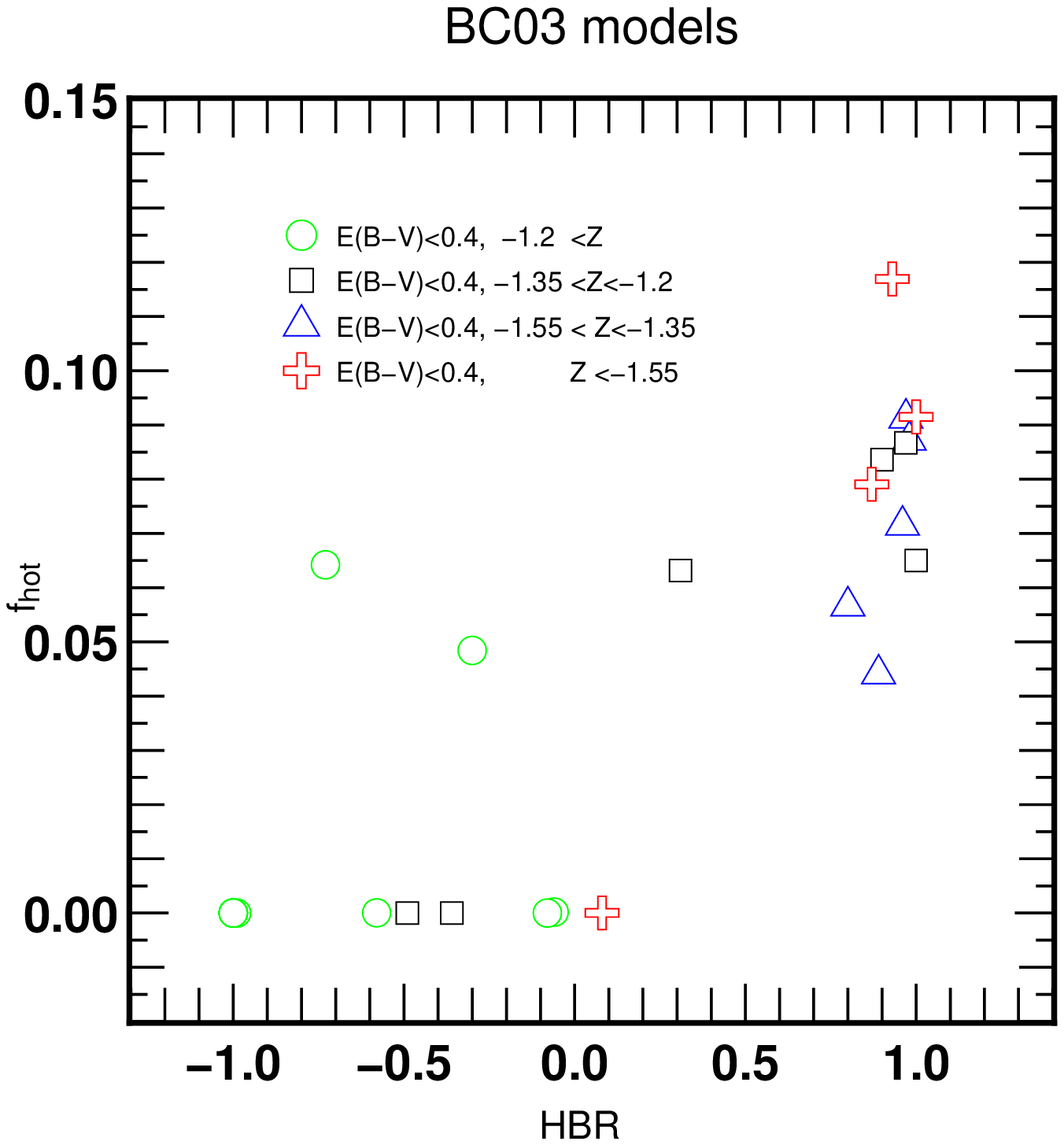}
\caption{Hot fraction as a function of horizontal branch morphology for the clean sample, using Vazdekis et al. ({\em left}), and BC03 models ({\em right}).}
\label{f:bc03}
\end{figure*}

The left panel of Fig. \ref{f:bc03} shows the results for the cleaned sample. The correlation between $\fy$ and blue HB morphology is now much clearer:
spurious bursts of recent star formation are very likely to appear in GCs with $\hbr > 0$, while $\fy=0$ for all the GCs with $\hbr < 0$. The intensity of the spurious young burst can be as high as 10\% of the total optical light of the cluster.
The spurious recent bursts correlate with HB morphology.

However, we find that 4 (out of 13) GCs with BHB have an almost $\fy\approx 0$. To check this last puzzling point and more generally the robustness of the $\fy$-HBR correlation we investigated the influence of the theoretical error in the SSP models.


\begin{table}
 \begin{tabular}{lccccc} 
\hline 
\\
NGC & E(B-V) & Z & HBR & ${\rm f_{hot}^{Vaz}}$ & ${\rm f_{hot}^{BC03}}$  \\
(1) & (2) & (3) & (4) & (5) & (6) 
\\
\hline
\hline
 104 & 0.04 & -0.7 & -0.99 & 0 & 0 \\ 
 1851 & 0.02 & -1.21 & -0.36 & 0 & 0 \\ 
 1904 & 0.01 & -1.55 & -0.89 & 0 & 8 \\ 
 2298 & 0.15 & -1.97 & 0.93 & 3 & 11 \\ 
 2808 & 0.23 & -1.29 & -0.49 & 0 & 0 \\ 
 3201 & 0.21 & -1.56 & 0.08 & 1 & 0 \\ 
 5286 & 0.24 & -1.51 & 0.8 & 0 & 5 \\ 
 5904 & 0.03 & -1.26 & 0.31 & 8 & 6 \\ 
 5927 & 0.45 & -0.64 & -1 & 0 & 0 \\ 
 5946 & 0.54 & -1.54 & 1 & 4 & 2 \\ 
 5986 & 0.27 & -1.53 & 0.97 & 0 & 9 \\ 
 6121 & 0.4 & -1.15 & -0.06 & 0 & 0 \\ 
 6171 & 0.33 & -1.13 & -0.73 & 0 & 6 \\ 
 6218 & 0.19 & -1.32 & 0.97 & 7 & 8 \\ 
 6235 & 0.36 & -1.36 & 0.89 & 3 & 4 \\ 
 6254 & 0.28 & -1.51 & 0.98 & 4 & 8 \\ 
 6266 & 0.47 & -1.2 & 0.32 & 0 & 0 \\ 
 6284 & 0.28 & -1.27 & 1 & 8 & 6 \\ 
 6304 & 0.52 & -0.66 & -1 & 0 & 0 \\ 
 6316 & 0.51 & -0.9 & -1 & 1 & 1 \\ 
 6333 & 0.35 & -1.65 & 0.87 & 0 & 7 \\ 
 6342 & 0.46 & -1.01 & -1 & 0 & 0 \\ 
 6352 & 0.19 & -0.7 & -1 & 0 & 0 \\ 
 6356 & 0.28 & -0.74 & -1 & 0 & 0 \\ 
 6362 & 0.08 & -1.17 & -0.58 & 0 & 0 \\ 
 6522 & 0.48 & -1.39 & 0.71 & 8 & 7 \\ 
 6528 & 0.62 & -0.1 & -1 & 6 & 3 \\ 
 6544 & 0.73 & -1.38 & 1 & 5 & 2 \\ 
 6553 & 0.8 & -0.2 & -1 & 13 & 10 \\ 
 6624 & 0.28 & -0.7 & -1 & 0 & 0 \\ 
 6626 & 0.38 & -1.21 & 0.9 & 10 & 8 \\ 
 6637 & 0.16 & -0.78 & -1 & 0 & 0 \\ 
 6638 & 0.4 & -1.08 & -0.3 & 0 & 4 \\ 
 6652 & 0.09 & -1.1 & -1 & 0 & 0 \\ 
 6723 & 0.05 & -1.14 & -0.08 & 0 & 0 \\ 
 6752 & 0.04 & -1.57 & 1 & 0 & 9 \\ 
 7089 & 0.06 & -1.49 & 0.96 & 4 & 7 \\ 
\hline
 \end{tabular}

\label{t:GCtab}
\caption{Properties of the sample of GCs  and results. (1): NGC name, (2): color excess, (3): Metalliciy {\rm [Fe/H]}, (4): horizontal branch ratio, (5) Fraction of light in the hot component in \%, computed from the SADs obtained with the Vazdekis models, (6): same as (5) with BC03.
Columns (1)-(4) are taken from \cite{schiavon05}, and for (5)-(6), see Sec. \ref{s:results}.}
\end{table}

\subsection{Model dependence}

We repeated the analysis of the sample using the BC03 models. 
The results are shown in the right panel of Fig. \ref{f:bc03}. First of all, the correlation between $\fy$ and HBR is recovered, and is therefore a robust result of our study.
However, the frequency of spurious young bursts is higher in general with this model, both in GCs with BHB and RHB. In particular, the low-Z GCs for which $\fy=0$ with the Vazdekis models, now all have a measurable $\fy > 0$. On the other hand, 2 of the red HB GCs, which previously had $\fy=0$ now have a non-zero hot fraction.





The overall consistency of the results obtained using the 2 models indicates that, for the regime studied here (sub-solar metallicity, age $>$ 6 Gyr), the systematic error introduced by the uncertainties in the modelling of the HB are larger than the other error sources (differences in the underlying stellar libraries, accuracy of the stellar parameters and coverage of the latter, interpolation procedures).

\subsection{Method dependence}
In \cite{koleva08}, the authors performed a similar experiment, however with a different, parametric method \citep{NBURSTS}, where the young SSP was replaced with a hot star spectrum. They also found a correlation between the HBR and the luminous weight of the hot star, although much weaker than the one presented in this paper in Fig. \ref{f:bc03}. 
In a different type of experiment, NBURSTS and STECKMAP were found to give consistent results for luminosity-weighted ages and metallicities. Moreover, ULYSS (Koleva et al., in prep.), another non-parametric code, was also found to give similar results to STECKMAP in the reconstruction of the star formation history of dwarf elliptical galaxies \citep{koleva09}. Because of this agreement, we suggest that the clearer $\fy$-$\hbr$ correlation we find here with respect to \cite{koleva08} is essentially the consequence of the wider wavelength range of the models we used. Indeed, the Vazdekis et al. models allow us to exploit the full coverage of the data into the blue, down to 3500 {\AA}, while \cite{koleva08} used PEGASE-HR \citep{PEG-HR}, which starts at 3900 {\AA}.
This gives us good confidence in the fact that our results are robust to a change of method, as well as a change of SSP models, provided that the wavelength range are kept similar.



\section{Discussion}
\label{s:discussion}
\subsection{The confusion zone}
We identified a range in metallicity (${\rm [Fe/H]}<-1.2$) where the analysis of the integrated spectra of stellar populations in the optical is highly likely ($70-100\%$ probability depending on the model) to lead to the detection of a spurious young burst of star formation, contributing up to 12\% of the light. 
The luminosity of this spurious burst correlates with HB morphology, and increases towards bluer HB. This result is independent of the SSP model and of the method used.

The consequence for studies of stellar populations of galaxies is that any detection of a young burst of star formation superimposed on an old metal-poor population (${\rm [Fe/H]}<-1.2$) from analysis of the optical spectrum is possibly spurious {\em if the contribution of the young populations is smaller than 12\% and no emission lines typical of HII regions are seen}. 
Moreover, since galactic stellar populations are composite in general, smaller young fractions can still be the signature of BHB stars even if the old component has a { luminosity-weighted} metallicity larger than $-1.2$. For example, following our results, an old composite population, with $50$\% of mass in solar metallicity stars and $50$\% of mass in ${\rm [Fe/H]}=-1.2$ stars could still display $5\%$ of the light in a fake young burst. What exactly does it mean for extragalactic studies ? In the following we discuss the implications of our result for spectroscopic studies of various objects in integrated light.

\subsection{Field and cluster HB stars}
So far we have considered exclusively GCs. It is not obvious that our result also applies to integrated spectroscopy of field stars, which is the method of choice for extragalactic studies.

Indeed, some stages of stellar evolution seem to differ depending on membership to a GC or to the field, such as metal-rich RR-Lyrae stars \citep{layden99,catelan06,matsunaga07}. 
However, these differences seem to disappear at low Z (i.e. the regime of this work). Moreover, differences in the properties between field and GCs HB stars (other than RR Lyrae) have not been detected. It is not clear why some stages are affected by GC membership while others are not. Two processes at least are expected to be significantly more important in cluster environments than in the field. These are binary stellar evolution (higher binarity in clusters), and binary disruption (decreasing binary frequency), as shown for example in \cite{guhathakurta98} and \cite{sollima07}. 
As a matter of fact, blue HB morphology is indeed observed in the field as well: the CMDs of {\cite{dolphin02}} show that the dwarf galaxies Leo I and especially Ursa Minor have a well defined BHB.
In the absence of strong contrary evidence, we consider that the HB of the GCs analysed here are representative of the HB morphologies of at least a fraction of the field stellar populations of galaxies.

\subsection{Metal-rich populations}
The highest metallicity object of our clean sample is NGC104, with {\rm [Fe/H]}=-0.7. We were unable to use the higher metallicity part of the original sample because it happens to be also the most extincted, and thus the most contaminated. However, upon visual inspection of the CMDs of NGC6528 and NGC6553 ({\rm [Fe/H]}=-0.1 and {\rm [Fe/H]}=-0.2), no evidence for a blue HB is found, and these populations should therefore be well behaved with respect to star formation history reconstructions. This extends our results to quasi-solar metallicities. At even higher metallicities, we recall that spectroscopy of elliptical galaxies in the optical suggests small but recent star formation \citep{trager00}, in the form of a ``frosting'' of young stars, reminiscent of the artifacts we have observed for GCs with BHB. This is a very important aspect since the halt of star formation in massive galaxies, as inferred from their red colours, motivates a large amount of studies in the search for quenching mechanisms, either through hot shocks and the end of cold streams \citep{BD03,keres05,ocvirk08}, or AGN feedback \citep{hopkins07,delucia07}. We note however that evidence for recent star formation in early-type galaxies is also found in IR data \citep{lyubenova08}, which is little sensitive to any EHB/BHB contribution.
To probe such high metallicities, an integrated spectrum of NGC6791, an old (8-9 Gyr), metal-rich (twice solar) open cluster, would be very useful. Interestingly, several extreme horizontal branch stars (EHB), even hotter than usual BHB stars, have been found in this object \citep{kaluzny92,Liebert94}, although without any sizeable BHB counterpart. However, it is unlikely that this small population of very hot evolved stars can produce such strong artifacts in SFH reconstructions as we have seen from optical light. 
Their contribution peaks in the FUV and affects the optical light much less than the BHB studied here. As an alternative to NGC6791, one could study old metal-rich extragalactic GCs, although we would then miss the accurate photometric determination of their HB morphology from a CMD.

\subsection{M32}
The rightmost panel of Fig. 6 of Coelho et al. 2009 shows that except for the nucleus, a large fraction (about $\approx 50$ \%) of the light of M32 is produced by a ${\rm [Fe/H]}<-1.2$ old stellar population. Our study shows that in this regime, a spurious young burst accounting for $\approx 5\%$ is indeed likely to appear.
However, in the nucleus, the underlying stellar population is metal-rich. Even though we do not constrain well the extent of the ``confusion zone'' in the high Z regime, the young component is stronger than the fake bursts we found: it accounts for 20 \% of the light (in a similar wavelength range) while we found maximum spurious contributions of 10-12\%.
Hence, our result supports the claims of Coelho et al. 2009, mainly that there has been a recent burst of star formation in the center of M32, while the young component seen in the outskirts is also compatible with being an artifact due to the blue HB morphology of the system.

\section{Conclusions}
The analysis of integrated spectra of galactic GCs \cite{schiavon05} with STECKMAP yields two families of stellar age distributions:
\begin{itemize}
\item{Type I: genuinely old populations, with no contribution from components younger than $10^{9.5}$ yr.}
\item{Type II: old populations with a superimposed hot component younger than $10^{8.5}$ yr contributing a fraction $\fy$ to the light of the GC.}
\end{itemize}
GCs with red HB and a type II SAD are subject to background/foreground contamination. Since the latter correlates with color excess E(B-V), we define a clean sample (with respect to contamination) by removing GCs with E(B-V)$>0.4$.
In this new sample, red HB GCs all have type I SADs, while blue HB GCs have mostly type II SADs, with up to 12\% of the light in the hot component.
This is the signature of inaccuracies in the modelling of the horizontal branch: in the model tracks used in Vazdekis et al. and \cite{BC03}, the blue HB is often too short compared to observations. As a consequence, the observed integrated spectrum contains an additional hot component with respect to the corresponding model SSP, which gives rise to the $\fy$-HBR correlation. This is the main result of the paper, and it is found with both SSP models Vazdekis et al. and BC03.

We have thus identified a ``confusion zone'', where spurious starbursts are very likely to appear in star formation history reconstructions, eventhough the observed population is genuinely old. These fake starbursts are young ($<10^{8.5}$ yr) and occur in 70-100\% of cases depending on the SSP model, and can weight up to 12$\%$ of the light in the optical. This confusion is driven by HB morphology and is thus mainly a low-metallicity effect (${\rm [Fe/H]}<-1.2$). In this regime, any minor young burst of star formation can be an artifact, { especially if no emission lines associated with star formation are seen.} We expect this prediction to be independent of the interpretation method and models, as long as the evolutionary tracks used for the spectral synthesis are not tuned specifically to account for BHB morphology. The exact weight allowed for the fake starburst depends on the wavelength range considered. It should increase for domains bluer than in this work, and decrease towards the red.

The highest metallicity of our clean sample is {\rm [Fe/H]}=-0.7, which prevents us from properly probing the high-Z regime. We plan to do so by studying old metal-rich extragalactic GCs or an integrated spectrum of NGC6791.

We wish to encourage the stellar evolution and extragalactic communities to consolidate the recent progresses made on HB morphology modeling by including the relevant tracks into future spectral synthesis models such as Vazdekis et al.. However we note that accurate fitting of blue HB morphology in GCs by fine tuning of He abundance dispersion such as in \cite{lee05} introduces additional  parameters, and possibly new degeneracies in spectrum interpretation. Moreover, proper use of these tracks requires a stellar library, empirical or theoretical, for a range of non-solar abundance ratios, reliable everywhere on the CMD. Such models are still in their infancy, despite improvements such as \cite{coelho07}.

This study is also an interesting sanity check for STECKMAP, which is found to behave very well. We propose that the exercise of analysing the E(B-V)-filtered sample and checking the results dependence on HBR as we have done be a test for all the star formation history reconstruction methods, such as MOPED \citep{moped01}, VESPA \citep{vespa07}, STARLIGHT \citep{CF04-1} or \cite{wild07}, and urge other authors to perform the same experiment.

Finally, the fact that STECKMAP  associates BHB morphology with a type II SAD with a high success rate, shows that HB morphology can be determined in integrated light through full spectrum fitting, and in the optical domain, as was realised by \cite{schiavon04}. 
This technique could thus become a valuable tool for extragalactic cluster studies: by removing the need for UV (i.e. space) observations and resolved populations (i.e. relatively nearby), it gives access to samples and environments otherwise unreachable.





\section*{Acknowledgements}
The author thanks P. Coelho, A. Nebot Gomez-Moran, A. Lan\c{c}on, D. Leborgne, K. Freeman, P. Prugniel and M. Koleva for useful comments, and M. Koleva again for providing accurate individual spectral masks for bad pixels and sky subtraction residuals. This work is supported by a grant from the Deutsche Luft und Raumfahrt (DLR).
It has made use of the IAC-STAR Synthetic CMD computation code based on \cite{aparicio04}. IAC-STAR is suported and maintained by the computer division of the Instituto de Astrof\'{i}sica de Canarias. The author thanks D.~Munro for freely distributing his Yorick programming language\footnote{http://www.maumae.net/yorick/doc/index.html}, and E.~Thi\'ebaut for sharing his Yorick optimization package \texttt{OptimPack}.

\bibliographystyle{mn2e}
\bibliography{mybib}

\begin{thebibliography}{}

\bibitem[\protect\citeauthoryear{{Anders} \& {Grevesse}}{{Anders} \&
  {Grevesse}}{1989}]{anders89}
{Anders} E.,  {Grevesse} N.,  1989, \gca, 53, 197

\bibitem[\protect\citeauthoryear{{Aparicio} \& {Gallart}}{{Aparicio} \&
  {Gallart}}{2004}]{aparicio04}
{Aparicio} A.,  {Gallart} C.,  2004, \aj, 128, 1465

\bibitem[\protect\citeauthoryear{{Beasley}, {Sharples}, {Bridges}, {Hanes},
  {Zepf}, {Ashman} \& {Geisler}}{{Beasley} et~al.}{2000}]{beasley2000}
{Beasley} M.~A.,  {Sharples} R.~M.,  {Bridges} T.~J.,  {Hanes} D.~A.,  {Zepf}
  S.~E.,  {Ashman} K.~M.,    {Geisler} D.,  2000, \mnras, 318, 1249

\bibitem[\protect\citeauthoryear{{Bedin}, {Piotto}, {Anderson}, {Cassisi},
  {King}, {Momany} \& {Carraro}}{{Bedin} et~al.}{2004}]{bedin04}
{Bedin} L.~R.,  {Piotto} G.,  {Anderson} J.,  {Cassisi} S.,  {King} I.~R.,
  {Momany} Y.,    {Carraro} G.,  2004, \apjl, 605, L125

\bibitem[\protect\citeauthoryear{{Bertelli}, {Bressan}, {Chiosi}, {Fagotto} \&
  {Nasi}}{{Bertelli} et~al.}{1994}]{bertelli94}
{Bertelli} G.,  {Bressan} A.,  {Chiosi} C.,  {Fagotto} F.,    {Nasi} E.,  1994,
  \aaps, 106, 275

\bibitem[\protect\citeauthoryear{{Birnboim} \& {Dekel}}{{Birnboim} \&
  {Dekel}}{2003}]{BD03}
{Birnboim} Y.,  {Dekel} A.,  2003, \mnras, 345, 349

\bibitem[\protect\citeauthoryear{{Bruzual} \& {Charlot}}{{Bruzual} \&
  {Charlot}}{2003}]{BC03}
{Bruzual} G.,  {Charlot} S.,  2003, \mnras, 344, 1000

\bibitem[\protect\citeauthoryear{{Catelan}, {Stetson}, {Pritzl}, {Smith},
  {Kinemuchi}, {Layden}, {Sweigart} \& {Rich}}{{Catelan}
  et~al.}{2006}]{catelan06}
{Catelan} M.,  {Stetson} P.~B.,  {Pritzl} B.~J.,  {Smith} H.~A.,  {Kinemuchi}
  K.,  {Layden} A.~C.,  {Sweigart} A.~V.,    {Rich} R.~M.,  2006, \apjl, 651,
  L133

\bibitem[\protect\citeauthoryear{{Chernoff} \& {Djorgovski}}{{Chernoff} \&
  {Djorgovski}}{1989}]{chernoff89}
{Chernoff} D.~F.,  {Djorgovski} S.,  1989, \apj, 339, 904

\bibitem[\protect\citeauthoryear{{Chilingarian}, {Prugniel}, {Sil'Chenko} \&
  {Koleva}}{{Chilingarian} et~al.}{2007}]{NBURSTS}
{Chilingarian} I.,  {Prugniel} P.,  {Sil'Chenko} O.,    {Koleva} M.,  2007, in
  {Vazdekis} A.,  {Peletier} R.~F.,  eds, IAU Symposium Vol.~241 of IAU
  Symposium, {NBursts: Simultaneous Extraction of Internal Kinematics and
  Parametrized SFH from Integrated Light Spectra}.
pp 175--176

\bibitem[\protect\citeauthoryear{Cid~Fernandes, Mateus, Sodre, Stasinska \&
  Gomes}{Cid~Fernandes et~al.}{2004}]{CF04-1}
Cid~Fernandes R.,  Mateus A.,  Sodre L.,  Stasinska G.,    Gomes J.~M.,  2004,
  astro-ph/0412481

\bibitem[\protect\citeauthoryear{{Coelho}, {Bruzual}, {Charlot}, {Weiss},
  {Barbuy} \& {Ferguson}}{{Coelho} et~al.}{2007}]{coelho07}
{Coelho} P.,  {Bruzual} G.,  {Charlot} S.,  {Weiss} A.,  {Barbuy} B.,
  {Ferguson} J.~W.,  2007, \mnras, 382, 498

\bibitem[\protect\citeauthoryear{{Coelho}, {Mendes de Oliveira} \& {Cid
  Fernandes}}{{Coelho} et~al.}{2009}]{coelho09}
{Coelho} P.,  {Mendes de Oliveira} C.,    {Cid Fernandes} R.,  2009, ArXiv
  e-prints

\bibitem[\protect\citeauthoryear{{Cox}}{{Cox}}{2000}]{allens}
{Cox} A.~N.,  2000, {Allen's astrophysical quantities}

\bibitem[\protect\citeauthoryear{{De Lucia} \& {Blaizot}}{{De Lucia} \&
  {Blaizot}}{2007}]{delucia07}
{De Lucia} G.,  {Blaizot} J.,  2007, \mnras, 375, 2

\bibitem[\protect\citeauthoryear{{Dolphin}}{{Dolphin}}{2002}]{dolphin02}
{Dolphin} A.~E.,  2002, \mnras, 332, 91

\bibitem[\protect\citeauthoryear{{Dotter}}{{Dotter}}{2008}]{dotter08}
{Dotter} A.,  2008, \apjl, 687, L21

\bibitem[\protect\citeauthoryear{{Ferraro}, {Origlia}, {Testa} \&
  {Maraston}}{{Ferraro} et~al.}{2004}]{ferraro04}
{Ferraro} F.~R.,  {Origlia} L.,  {Testa} V.,    {Maraston} C.,  2004, \apj,
  608, 772

\bibitem[\protect\citeauthoryear{{Girardi}, {Bressan}, {Bertelli} \&
  {Chiosi}}{{Girardi} et~al.}{2000}]{girardi00}
{Girardi} L.,  {Bressan} A.,  {Bertelli} G.,    {Chiosi} C.,  2000, \aaps, 141,
  371

\bibitem[\protect\citeauthoryear{{Guhathakurta}, {Webster}, {Yanny},
  {Schneider} \& {Bahcall}}{{Guhathakurta} et~al.}{1998}]{guhathakurta98}
{Guhathakurta} P.,  {Webster} Z.~T.,  {Yanny} B.,  {Schneider} D.~P.,
  {Bahcall} J.~N.,  1998, \aj, 116, 1757

\bibitem[\protect\citeauthoryear{{Han}, {Podsiadlowski} \& {Lynas-Gray}}{{Han}
  et~al.}{2007}]{han07}
{Han} Z.,  {Podsiadlowski} P.,    {Lynas-Gray} A.~E.,  2007, \mnras, 380, 1098

\bibitem[\protect\citeauthoryear{{Heavens}, {Panter}, {Jimenez} \&
  {Dunlop}}{{Heavens} et~al.}{2004}]{heavensnature}
{Heavens} A.,  {Panter} B.,  {Jimenez} R.,    {Dunlop} J.,  2004, \nat, 428,
  625

\bibitem[\protect\citeauthoryear{{Hinshaw}, {Weiland}, {Hill}, {Odegard},
  {Larson}, {Bennett}, {Dunkley}, {Gold}, {Greason} \& {Jarosik}}{{Hinshaw}
  et~al.}{2009}]{WMAP5}
{Hinshaw} G.,  {Weiland} J.~L.,  {Hill} R.~S.,  {Odegard} N.,  {Larson} D.,
  {Bennett} C.~L.,  {Dunkley} J.,  {Gold} B.,  {Greason} M.~R.,    {Jarosik}
  2009, \apjs, 180, 225

\bibitem[\protect\citeauthoryear{{Hopkins}, {Bundy}, {Hernquist} \&
  {Ellis}}{{Hopkins} et~al.}{2007}]{hopkins07}
{Hopkins} P.~F.,  {Bundy} K.,  {Hernquist} L.,    {Ellis} R.~S.,  2007, \apj,
  659, 976

\bibitem[\protect\citeauthoryear{{Kaluzny} \& {Udalski}}{{Kaluzny} \&
  {Udalski}}{1992}]{kaluzny92}
{Kaluzny} J.,  {Udalski} A.,  1992, Acta Astronomica, 42, 29

\bibitem[\protect\citeauthoryear{{Kaviraj}, {Schawinski}, {Devriendt},
  {Ferreras}, {Khochfar}, {Yoon}, {Yi}, {Deharveng} \& {Boselli}}{{Kaviraj}
  et~al.}{2007}]{kaviraj07}
{Kaviraj} S.,  {Schawinski} K.,  {Devriendt} J.~E.~G.,  {Ferreras} I.,
  {Khochfar} S.,  {Yoon} S.-J.,  {Yi} S.~K.,  {Deharveng} J.-M.,    {Boselli}
  A.,  2007, \apjs, 173, 619

\bibitem[\protect\citeauthoryear{{Kere{\v s}}, {Katz}, {Weinberg} \&
  {Dav{\'e}}}{{Kere{\v s}} et~al.}{2005}]{keres05}
{Kere{\v s}} D.,  {Katz} N.,  {Weinberg} D.~H.,    {Dav{\'e}} R.,  2005,
  \mnras, 363, 2

\bibitem[\protect\citeauthoryear{{Koleva}, {De Rijcke}, {Prugniel}, {Zeilinger}
  \& {Michielsen}}{{Koleva} et~al.}{2009}]{koleva09}
{Koleva} M.,  {De Rijcke} S.,  {Prugniel} P.,  {Zeilinger} W.~W.,
  {Michielsen} D.,  2009, ArXiv e-prints

\bibitem[\protect\citeauthoryear{{Koleva}, {Prugniel}, {Ocvirk}, {Le Borgne} \&
  {Soubiran}}{{Koleva} et~al.}{2008}]{koleva08}
{Koleva} M.,  {Prugniel} P.,  {Ocvirk} P.,  {Le Borgne} D.,    {Soubiran} C.,
  2008, \mnras, 385, 1998

\bibitem[\protect\citeauthoryear{{Lan{\c c}on} \& {Mouhcine}}{{Lan{\c c}on} \&
  {Mouhcine}}{2000}]{lancon2000}
{Lan{\c c}on} A.,  {Mouhcine} M.,  2000, in {Lan{\c c}on} A.,  {Boily} C.~M.,
  eds, Massive Stellar Clusters Vol.~211 of Astronomical Society of the Pacific
  Conference Series, {Stochastic Fluctuations in the Spectrophotometric
  Properties of Star Clusters}.
pp 34--+

\bibitem[\protect\citeauthoryear{{Layden}, {Ritter}, {Welch} \&
  {Webb}}{{Layden} et~al.}{1999}]{layden99}
{Layden} A.~C.,  {Ritter} L.~A.,  {Welch} D.~L.,    {Webb} T.~M.~A.,  1999,
  \aj, 117, 1313

\bibitem[\protect\citeauthoryear{{Le Borgne}, {Rocca-Volmerange}, {Prugniel},
  {Lan{\c c}on}, {Fioc} \& {Soubiran}}{{Le Borgne} et~al.}{2004}]{PEG-HR}
{Le Borgne} D.,  {Rocca-Volmerange} B.,  {Prugniel} P.,  {Lan{\c c}on} A.,
  {Fioc} M.,    {Soubiran} C.,  2004, \aap, 425, 881

\bibitem[\protect\citeauthoryear{{Le Borgne}, {Bruzual}, {Pell{\'o}}, {Lan{\c
  c}on}, {Rocca-Volmerange}, {Sanahuja}, {Schaerer}, {Soubiran} \&
  {V{\'{\i}}lchez-G{\'o}mez}}{{Le Borgne} et~al.}{2003}]{stelib}
{Le Borgne} J.-F.,  {Bruzual} G.,  {Pell{\'o}} R.,  {Lan{\c c}on} A.,
  {Rocca-Volmerange} B.,  {Sanahuja} B.,  {Schaerer} D.,  {Soubiran} C.,
  {V{\'{\i}}lchez-G{\'o}mez} R.,  2003, \aap, 402, 433

\bibitem[\protect\citeauthoryear{{Lee}, {Demarque} \& {Zinn}}{{Lee}
  et~al.}{1994}]{lee94}
{Lee} Y.-W.,  {Demarque} P.,    {Zinn} R.,  1994, \apj, 423, 248

\bibitem[\protect\citeauthoryear{{Lee}, {Joo}, {Han}, {Chung}, {Ree}, {Sohn},
  {Kim}, {Yoon}, {Yi} \& {Demarque}}{{Lee} et~al.}{2005}]{lee05}
{Lee} Y.-W.,  {Joo} S.-J.,  {Han} S.-I.,  {Chung} C.,  {Ree} C.~H.,  {Sohn}
  Y.-J.,  {Kim} Y.-C.,  {Yoon} S.-J.,  {Yi} S.~K.,    {Demarque} P.,  2005,
  \apjl, 621, L57

\bibitem[\protect\citeauthoryear{{Liebert}, {Saffer} \& {Green}}{{Liebert}
  et~al.}{1994}]{Liebert94}
{Liebert} J.,  {Saffer} R.~A.,    {Green} E.~M.,  1994, \aj, 107, 1408

\bibitem[\protect\citeauthoryear{{Lyubenova}, {Kuntschner} \&
  {Silva}}{{Lyubenova} et~al.}{2008}]{lyubenova08}
{Lyubenova} M.,  {Kuntschner} H.,    {Silva} D.~R.,  2008, \aap, 485, 425

\bibitem[\protect\citeauthoryear{{Matsunaga}}{{Matsunaga}}{2007}]{matsunaga07}
{Matsunaga} N.,  2007, in {Kerschbaum} F.,  {Charbonnel} C.,   {Wing} R.~F.,
  eds, Why Galaxies Care About AGB Stars: Their Importance as Actors and Probes
  Vol.~378 of Astronomical Society of the Pacific Conference Series, {The
  Period--Luminosity Relation of Mira Variables in NGC 6388 and NGC 6441}.
pp 86--+

\bibitem[\protect\citeauthoryear{{Michielsen}, {Boselli}, {Conselice},
  {Toloba}, {Whiley}, {Arag{\'o}n-Salamanca}, {Balcells}, {Cardiel}, {Cenarro},
  {Gorgas}, {Peletier} \& {Vazdekis}}{{Michielsen} et~al.}{2008}]{michielsen08}
{Michielsen} D.,  {Boselli} A.,  {Conselice} C.~J.,  {Toloba} E.,  {Whiley}
  I.~M.,  {Arag{\'o}n-Salamanca} A.,  {Balcells} M.,  {Cardiel} N.,  {Cenarro}
  A.~J.,  {Gorgas} J.,  {Peletier} R.~F.,    {Vazdekis} A.,  2008, \mnras, 385,
  1374

\bibitem[\protect\citeauthoryear{{Norris}}{{Norris}}{2004}]{norris04}
{Norris} J.~E.,  2004, \apjl, 612, L25

\bibitem[\protect\citeauthoryear{{Ocvirk}, {Pichon}, {Lan{\c c}on} \&
  {Thi{\'e}baut}}{{Ocvirk} et~al.}{2006a}]{STECKMAP}
{Ocvirk} P.,  {Pichon} C.,  {Lan{\c c}on} A.,    {Thi{\'e}baut} E.,  2006a,
  \mnras, 365, 74

\bibitem[\protect\citeauthoryear{{Ocvirk}, {Pichon}, {Lan{\c c}on} \&
  {Thi{\'e}baut}}{{Ocvirk} et~al.}{2006b}]{STECMAP}
{Ocvirk} P.,  {Pichon} C.,  {Lan{\c c}on} A.,    {Thi{\'e}baut} E.,  2006b,
  \mnras, 365, 46

\bibitem[\protect\citeauthoryear{{Ocvirk}, {Pichon} \& {Teyssier}}{{Ocvirk}
  et~al.}{2008}]{ocvirk08}
{Ocvirk} P.,  {Pichon} C.,    {Teyssier} R.,  2008, \mnras, 390, 1326

\bibitem[\protect\citeauthoryear{{Panter}, {Heavens} \& {Jimenez}}{{Panter}
  et~al.}{2003}]{panter1}
{Panter} B.,  {Heavens} A.~F.,    {Jimenez} R.,  2003, \mnras, 343, 1145

\bibitem[\protect\citeauthoryear{{Prugniel}, {Soubiran}, {Koleva} \& {Le
  Borgne}}{{Prugniel} et~al.}{2007}]{elodie31}
{Prugniel} P.,  {Soubiran} C.,  {Koleva} M.,    {Le Borgne} D.,  2007, ArXiv
  Astrophysics e-prints

\bibitem[\protect\citeauthoryear{{Puzia}, {Kissler-Patig}, {Thomas},
  {Maraston}, {Saglia}, {Bender}, {Goudfrooij} \& {Hempel}}{{Puzia}
  et~al.}{2005}]{puzia05}
{Puzia} T.~H.,  {Kissler-Patig} M.,  {Thomas} D.,  {Maraston} C.,  {Saglia}
  R.~P.,  {Bender} R.,  {Goudfrooij} P.,    {Hempel} M.,  2005, \aap, 439, 997

\bibitem[\protect\citeauthoryear{{Reichardt}, {Jimenez} \&
  {Heavens}}{{Reichardt} et~al.}{2001}]{moped01}
{Reichardt} C.,  {Jimenez} R.,    {Heavens} A.~F.,  2001, \mnras, 327, 849

\bibitem[\protect\citeauthoryear{{S{\'a}nchez-Bl{\'a}zquez}, {Peletier},
  {Jim{\'e}nez-Vicente}, {Cardiel}, {Cenarro}, {Falc{\'o}n-Barroso}, {Gorgas},
  {Selam} \& {Vazdekis}}{{S{\'a}nchez-Bl{\'a}zquez} et~al.}{2006}]{MILES}
{S{\'a}nchez-Bl{\'a}zquez} P.,  {Peletier} R.~F.,  {Jim{\'e}nez-Vicente} J.,
  {Cardiel} N.,  {Cenarro} A.~J.,  {Falc{\'o}n-Barroso} J.,  {Gorgas} J.,
  {Selam} S.,    {Vazdekis} A.,  2006, \mnras, 371, 703

\bibitem[\protect\citeauthoryear{{Schawinski}, {Kaviraj}, {Khochfar}, {Yoon},
  {Yi}, {Deharveng}, {Boselli}, {Barlow} \& {Conrow}.}{{Schawinski}
  et~al.}{2007}]{schawinski07kav}
{Schawinski} K.,  {Kaviraj} S.,  {Khochfar} S.,  {Yoon} S.-J.,  {Yi} S.~K.,
  {Deharveng} J.-M.,  {Boselli} A.,  {Barlow} T.,    {Conrow}. 2007, \apjs,
  173, 512

\bibitem[\protect\citeauthoryear{{Schiavon}, {Rose}, {Courteau} \&
  {MacArthur}}{{Schiavon} et~al.}{2004}]{schiavon04}
{Schiavon} R.~P.,  {Rose} J.~A.,  {Courteau} S.,    {MacArthur} L.~A.,  2004,
  \apjl, 608, L33

\bibitem[\protect\citeauthoryear{{Schiavon}, {Rose}, {Courteau} \&
  {MacArthur}}{{Schiavon} et~al.}{2005}]{schiavon05}
{Schiavon} R.~P.,  {Rose} J.~A.,  {Courteau} S.,    {MacArthur} L.~A.,  2005,
  \apjs, 160, 163

\bibitem[\protect\citeauthoryear{{Sharina} \& {Davoust}}{{Sharina} \&
  {Davoust}}{2009}]{sharina09}
{Sharina} M.,  {Davoust} E.,  2009, \aap, 497, 65

\bibitem[\protect\citeauthoryear{{Smith} \& {Strader}}{{Smith} \&
  {Strader}}{2007}]{smith07}
{Smith} G.~H.,  {Strader} J.,  2007, Astronomische Nachrichten, 328, 107

\bibitem[\protect\citeauthoryear{{Sollima}, {Beccari}, {Ferraro}, {Fusi Pecci}
  \& {Sarajedini}}{{Sollima} et~al.}{2007}]{sollima07}
{Sollima} A.,  {Beccari} G.,  {Ferraro} F.~R.,  {Fusi Pecci} F.,
  {Sarajedini} A.,  2007, \mnras, 380, 781

\bibitem[\protect\citeauthoryear{{Tojeiro}, {Heavens}, {Jimenez} \&
  {Panter}}{{Tojeiro} et~al.}{2007}]{vespa07}
{Tojeiro} R.,  {Heavens} A.~F.,  {Jimenez} R.,    {Panter} B.,  2007, \mnras,
  381, 1252

\bibitem[\protect\citeauthoryear{{Tojeiro}, {Wilkins}, {Heavens}, {Panter} \&
  {Jimenez}}{{Tojeiro} et~al.}{2009}]{vespa09}
{Tojeiro} R.,  {Wilkins} S.,  {Heavens} A.~F.,  {Panter} B.,    {Jimenez} R.,
  2009, ArXiv e-prints

\bibitem[\protect\citeauthoryear{{Trager}, {Faber}, {Worthey} \&
  {Gonz{\'a}lez}}{{Trager} et~al.}{2000}]{trager00}
{Trager} S.~C.,  {Faber} S.~M.,  {Worthey} G.,    {Gonz{\'a}lez} J.~J.,  2000,
  \aj, 120, 165

\bibitem[\protect\citeauthoryear{{Wild}, {Kauffmann}, {Heckman}, {Charlot},
  {Lemson}, {Brinchmann}, {Reichard} \& {Pasquali}}{{Wild}
  et~al.}{2007}]{wild07}
{Wild} V.,  {Kauffmann} G.,  {Heckman} T.,  {Charlot} S.,  {Lemson} G.,
  {Brinchmann} J.,  {Reichard} T.,    {Pasquali} A.,  2007, \mnras, 381, 543

\bibitem[\protect\citeauthoryear{{Yi}}{{Yi}}{2008}]{Yi2008}
{Yi} S.~K.,  2008, in {Heber} U.,  {Jeffery} C.~S.,   {Napiwotzki} R.,  eds,
  Hot Subdwarf Stars and Related Objects Vol.~392 of Astronomical Society of
  the Pacific Conference Series, {The Current Understanding on the UV Upturn}.
pp~3--+

\bibitem[\protect\citeauthoryear{{Zoccali}, {Renzini}, {Ortolani}, {Bica} \&
  {Barbuy}}{{Zoccali} et~al.}{2001}]{zoccali01}
{Zoccali} M.,  {Renzini} A.,  {Ortolani} S.,  {Bica} E.,    {Barbuy} B.,  2001,
  \aj, 121, 2638

\end{thebibliography}
%

\label{lastpage}
\end{document}